\documentstyle[12pt]{article}

\def\be {\begin{equation}}
\def\ee {\end{equation}}
\newtheorem{theorem}{Theorem }
\begin{document}
\begin{titlepage}
\today          \hfill
\begin{center}
\hfill   Bulletin board quant-ph/9506014 \\

\vskip .2in

{\large \bf Quantum Mechanics with Event Dynamics.}
\footnote{Extended version of the talk given by the second author at the
XXVII-th Symposium of Reports on Mathematical Physics, Torun, 6-9 December,
1995}
\vskip .50in


\vskip .2in

Ph.~Blanchard${}^\flat$\footnote{
e-mail: blanchard@physik.uni-bielefeld.de}
\ and\ A.~Jadczyk${}^\sharp$\footnote{
e-mail: ajad@ift.uni.wroc.pl}

{\em ${}^\flat$ Faculty of Physics and BiBoS,
University of Bielefeld\\
Universit\"atstr. 25,
D-33615 Bielefeld\\
${}^\sharp$ Institute of Theoretical Physics,
University of Wroc{\l}aw\\
Pl. Maxa Borna 9,
PL-50 204 Wroc{\l}aw}
\end{center}

\vskip .2in

\begin{abstract}
Event generating algorithm corresponding to a linear master equation
of Lindblad's type is described and illustrated on two examples:
that of a particle detector and of a fuzzy clock. Relation to other
approaches to foundations of quantum theory and to description of quantum
measurements is briefly discussed.
\end{abstract}

\end{titlepage}
\newpage
\section{Introduction}
In a recent series of papers (cf \cite{blaja95a} and references therein)
we enhanced the standard framework of quantum mechanics endowing it with
event dynamics. In this extension, which will denote EEQT (for
Event Enhanced Quantum Theory), we go beyond the Schr\"odinger continuous
time evolution of wave packets - we also propose a class of algorithms
generating discrete events. From master equation that describes continuous
evolution of ensembles of coupled quantum+classical systems we derive a
unique piecewise deterministic random process that provides a stochastic
algorithm for generating sample histories of individual systems.
In the prsent contribution we will describe the essence of our approach. But
first we make a few comments on similarities and differences between EEQT and
several other approaches.

1) {\em The Standard Approach}\\
In the standard approach classical concepts are static. They are introduced
via measurement postulates developed by the founders of Quantum Theory.
But \lq measurement\rq\ itself is never precisely defined in the standard
approach and therefore measurement postulates can not be derived from the
formalism. One is supposed to believe Born's statistical interpretation
simply \lq because it works\lq\ . The standard interpretation alone does not
tell us what happens when a quantum system is under a continuous observation
(which, in fact, is always the case).

2) {\em Master Equation Dynamics and Continuous Observation Theory}\\
Continuous observation theory is usually based on successive applications
of the projection postulate. Each application of the projection
postulate maps pure states into mixed states. Thus repeated application
of the postulate leads to a master equation for a density matrix. Replacing
Schr\"odinger's dynamics by a master equation is also popular in quantum
optics (cf. \cite{carm93}) and in several attempts to reconcile quantum theory
with gravity (for a recent account see \cite{nan95}. In all these approaches
the authors usually believe that no classical system is introduced. All is
pure quantum. That is however not true. What is true is just the converse:
the largest possible classical system is introduced, but because it is
so large and so close to the eye - it easily escapes our sight. It is
assumed, without any justification, that jumps of quantum state vectors are
directly observable (whatever it means). These jumps are supposed to
constitute the only classical events. The weak point of this approach
is in the fact that going from the master equation, that describes statistical
ensembles, to a stochastic algorithm generating sample histories of
an individual system is non-unique. There are infinitely many
random processes that lead to the same master equation after averaging.
One can use diffusion stochastic differential equations or jump
processes, one can shift pieces of dynamics between Hamiltonian
evolution and collapse events.\\
The reason for this non--uniqueness is simple: there are infinitely many
mixtures that lead to the same density matrix. Diosi \cite{dio1,dio2} invented
a clever mathematical procedure for constructing a special \lq ortho
process\rq\ . It provides a definite algorithm in special cases of finite
degeneracy. It does not however remove non-uniqueness and also there is no
reason why nature should have chosen his special prescription causing quantum
state vector always to make the least probable transition: to one of the
orthogonal states.

3) {\em Bohmian Mechanics, Local Beables, Stochastic Mechanics}\\
In these approaches (cf. references \cite{bohm-un,bell-be,nelson85}) {\em
there is}\ an explicit classical system. Quantum state vector knows nothing
about this classical system. It evolves according to the unmodified
Schr\"odinger's dynamics. It acts on the classical system affecting the
classical dynamics (which is either causal or stochastic) without itself
being acted upon. There is a mysterious {\em quantum potential}\ : action
without re-action. All such schemes are inconsistent with quantum mechanics.
They can be shown to contradict indistinguishability of quantum mixtures that
are described by the same density matrix \cite{jad95a}. That it must be so
follows from quite general no--go theorems - cf. \cite{land91,ozawa92,ja94a}.
The fact that the above schemes allow us to distinguish between mixtures that
standard quantum mechanics consider indistinguishable need not be a weakness.
In fact, it may be an advantage because it may lead us beyond quantum
theory, it can provide us with means of faster than light communication -
provided experiment confirms this feature.

How does our approach compare to those above? First of all, as for today,
our approach is explicitly phenomenological. That is not to say that,
for instance, the standard approach is not phenomenological. Also
in the standard approach {\em we}\ must decide where do we finish
our quantum description and what do we \lq measure.\rq\ That does not
follow from the theory - it must be imputed from outside. However
we have been so much indoctrinated by Bohr's philosophy and its
apparent victory over Einstein's \lq realistic\rq\ dreams, and we are today
so used to this procedure, that we do not feel uneasiness here any more.
Somehow we believe that the future \lq quantum theory of everything\
will explain all events that happen. But chances are that this theory of
everything will explain nothing. It will be a dead theory. It will not even
have a Hamiltonian, because there will be no time. It will be a theory of
the world in which nothing happens by itself. It will answer {\em our}\
questions about certain {\em probabilities.} -- but it will not explain
why anything {\em happens}. \\
Our theory of event dynamics starts with an explicit
phenomenological split between a quantum system, which is not directly
observable, and a classical system where events happen that can be observed
and that are to be described and explained. In other words our starting
point is an explicit mathematical formulation of the Heisenberg's cut. The
quantum system may be as big as one wishes it to be, the classical system
may retreat more and more,
moved as far as we wish -- towards our sense organs, towards our brains,
towards our mental processes. But the further we retreat the less
{\em facts}\ we explain. At the extreme limit we will be able to explain
nothing but changes of our mental states i.e. only mental events. That state
of affairs may be considered satisfactory for those who adhere to idealistic
or eastern philosophies, but it need not be the one that enriches our
understanding of the true workings of nature. Probably, for most of
practical purposes, it is sufficient to retreat with the quanto--classical cut
till the photon detection processes, which can be treated as the primitive
events. However, our event mechanics works quite well when the cut between
the quantum and the classical is expressed in engineering language: like
quantum SQUID coupled to classical radio-frequency circuit, or quantum
particle coupled to its position detector, for instance to a cloud chamber.
Once the split between the quantum and the classical is fixed, then
the coupling between both system is described in terms of a special
master equation. Because of its special form there is a unique
random process in the space of pure states of the total system that
reproduces this master equation. The process gives an algorithm for
generating sample histories. It is of piecewise deterministic
character. It consists of periods of continuous evolution interrupted
by jumps and events that happen at random times. The continuous
evolution of the quantum system is described by a -- modified by the coupling
-- non--unitary Schr\"odinger's equation. The jump times have a Poissonian
character, with their jump rates dependent on the actual state of both:
quantum and classical system. The back action of the classical system on the
quantum one shows up in two ways: first of all  by modifying the Schr\"odinger
evolution between jumps by a non--unitary damping, second by causing quantum
state to jump at event times. Notice that the master equation describing
statistical properties is linear, while evolution of individual system
is non--linear. This agrees with Turing's aphorism stating that
\lq prediction must be linear, description must be non--linear\rq\
\cite{hodgesa}.

Our theory, even if it works well and if it has a practical
value, should be considered not as a final scheme of things, but merely as
a step that may help us to find a description of nature that is
more satisfactory than the one proposed by the orthodox quantum
philosophy. Pure quantum theory proposes a universe that is dead -
nothing happens, nothing is real - apart of questions asked by
mysterious \lq observers\rq\ . Our theory of event mechanics, described here,
makes the universe \lq running\rq\ again. It has gotten however arrow of time
that is driven by a fuzzy quantum clock. It also needs a roulette. This is
hard to accept for most of us. We would like to believe that nature is ruled
by a perfect order. Even if we do not share Einstein's dissatisfaction with
quantum theory, we tend to understand his disgust  at the very thought of God
playing dice. On the other hand using probability theory may be the only
way of describing in finite terms the universe that has an infinite
complexity.
It may  be that we will never
know the ultimate secret, nevertheless the mechanism proposed by EEQT brings
a hope of restoring some order that we are seeking.
Namely, the quanto--classical clock that we describe below works by
itself. It is true that it needs a roulette but the
roulette is a {\em classical}\ roulette. We need
only {\em classical probability,}\ and classical random processes. That
is some progress, because nowadays we know more about complexity theory,
theory of random sequences, and theory of chaotic phenomena. Each year we
find new ways of generating apparently random phenomena out of deterministic
algorithms of sufficient complexity. In fact, our event generating algorithm
is successfully simulated with a completely deterministic computer.
The crucial problem here is the necessary computing power. Moreover, the
algorithm is non--local. We do not know how nature manages to make its world
clock running with no or little effort.
We must yet learn it.

\section{The Event Engine}
We will describe in this section the event producing algorithm that
results from our theory. The algorithm is simple, the master equation
that it leads to is also easy to write down. What is more difficult
is proving that the correspondence between statistical description
provided by the master equation implies the event algorithm uniquely
-- cf Ref. \cite{blaja95a}.
To make the idea as clear as possible we will assume that our classical
system admits only finite number of states. We will call these states
$\alpha=1,\ldots ,m$. There are $m^2-1$ possible events - labeled by
pairs $\alpha\neq\beta$. For each $\alpha$ let ${\cal H}_\alpha$ be the
Hilbert space of the quantum system. Usually all these Hilbert spaces
are isomorphic or even identical. But it costs us nothing to allow for
a more general setting, so that the transition $\alpha\rightarrow\beta$
may correspond to phase transition, where Hilbert space must also change.
We then need $m^2$ operators (or $m^2-m$ operators in a symmetric case - see
below): $m$ Hermitian operators $H_\alpha$ -
the Hamiltonians $H_\alpha :{\cal H}\rightarrow{\cal H}_\alpha$, and
$m^2-m$ operators $g_{\alpha\beta}:{\cal H}_\beta\rightarrow{\cal H}_\alpha$.
Thus our operator valued matrix $g_{\alpha\beta}$ has zeros on the diagonal.
The theory becomes most symmetric if a so called \lq detailed balance
condition\rq\, is satisfied, that is if
$g_{\alpha\beta}^\star=g_{\beta\alpha}$.
But working models may be produced without imposing this kind of symmetry
(for instance, our two examples in the next section are not symmetric).
The operators $H_\alpha, g_{\alpha\beta}$ may depend explicitly on time.
We will not make this dependence explicit but all our formulas below
are written in such a way that they remain valid in this more general case.\\
Before describing our event generating process let us introduce
a convenient notation; for any $\psi_\alpha\in{\cal H}_\alpha$ denote
\be
\Lambda_\alpha=\sum_\beta g_{\beta\alpha}^\star g_{\beta\alpha},
\ee
\be
\lambda_\alpha(\psi_\alpha)=(\psi_\alpha,\Lambda_\alpha \psi_\alpha),
\ee
\be
p_\beta(\psi_\alpha)=\frac{\Vert g_{\beta\alpha} \psi_\alpha \Vert^2}
{\lambda_\alpha (\psi_\alpha)}.
\ee
\subsection{Event generation}
The algorithm powering our event engine is described
by following the steps 1)--6) below.\\
\vskip 0.5cm
\noindent
$1$) Suppose at time $t=t_0$ classical system is in a state $\alpha$ and
quantum system is in a state $\psi_\alpha (t_0)\in{\cal H}_\alpha$.\\
$2$) Choose uniform random number $r\in (0,1)$.\\
$3$) Propagate $\psi_\alpha(t_0)$ in ${\cal H}_\alpha$ forward in time by
solving:
\be
{\dot \psi}_\alpha =
(-iH_\alpha - {1\over 2}\Lambda_\alpha){\psi}_\alpha
\ee
until $t=t_1$, where $t_1$ is defined by
\be
\Vert \psi_\alpha(t_1) \Vert^2=r
\label{eq:time} \ee
$4$) Choose uniform random number $r_1\in (0,1)$\\
$5$) Run through
the classical states $\beta=1,2,\ldots ,m$ until you reach $\beta=\alpha_1$
for which
\be
\sum_{\beta=1}^{\alpha_1} p_\beta(\psi_\alpha(t_1))\geq r_1 .
\ee
$6$)
Goto $1$) replacing $t_0\rightarrow t_1$, $\alpha\rightarrow\alpha_1$
and
$\psi_\alpha(t_0)\rightarrow g_{\alpha_1\alpha}\psi_\alpha (t_1)/
\Vert g_{\alpha_1\alpha}\psi_\alpha (t_1) \Vert $.\\

\noindent
{\bf Remark 1} According to the theory developed in Ref. \cite{blaja95a}
the jump process is an inhomogeneous Poisson process with intensity
function $\lambda_\alpha (t)$. One way to simulate such a process is to
move forward in time by small time intervals $\Delta t$, and make
independent decisions for jumping with probability $\lambda_\alpha (t)
\Delta t$. This leads to the probability $p$ of a jump to occur in the
time interval $(t_0,t)$ given by:
\be
p=1-\exp (-\int_{t_0}^t \lambda_\alpha (s) ds).\ee
By using the identity
$\log f(t) -\log f (t_0) = \int_{t_0}^t {\dot f (s)}/f(s)\; ds$ it
is easy to see that $p=1-\Vert \psi_\alpha(t) \Vert^2$ --
which simplifies simulation -- as we did in steps 2),3) above. This
observation throws also some new light on those approaches to
quantum mechanical description of particle decays that were based
on non-unitary evolution.\\
{\bf Remark 2} The algorithm above involves playing a roulette. If nature
is using this algorithm running her event engine, then the timing of
each next event is decided beforehand in step 2). But even if $r$ is
already chosen, still there is a possibility to delay or to hasten the
next event provided one has the ability to manipulate the time-dependence of
$g_{\beta\alpha}(t)$ that enter $\Lambda_\alpha (t)$ in Eq. (\ref{eq:time}).
\subsection{Master equation}
By repeating the above event generating algorithm many times, or by observing
time series of events for a prolonged time, we will notice certain
regularities and certain statistical tendencies. There are many ways
of collecting data that we consider of interest. For instance, we may
ask what is the average time necessary for arriving at a particular
classical state or a succession of states. But we can also ask more
standard question: suppose we repeat our simulation many times always
starting with the same state at the same initial time $t_0$, and ending it at
the same final time $t$. Then we will arrive at different final states with
different probabilities. Let $\alpha,\psi_\alpha,t_0$ be the initial
state, and let $\mu(\alpha,\psi_\alpha,t_0;\beta,\psi_\beta,t)$ be the
probability density of arriving at the state $(\beta,\psi_\beta)$ at
time $t$. We may associate with this probability distribution a family
of density matrices:
\be\rho_\beta =  \int
\mu(\alpha,\psi_\alpha,t_0;\beta,\psi_\beta,t) |\psi_\beta><\psi_\beta |
d \psi_\beta ,
\ee
so that $\sum_\beta {\mbox Tr} \rho_\beta =1$. This association is many
to one. We lose this way information. Nevertheless, as shown in
\cite{blaja95a,jakol}, the following theorem
holds:
\begin{theorem}
The family $\rho_\beta (t)$ satisfies linear differential equation
\be
{\dot \rho}_\beta =-i[H_\beta,\rho_\beta]+\sum_{\gamma\neq\beta}
g_{\beta\gamma}
\rho_\gamma g_{\beta\gamma}^\star-{1\over2}\{\Lambda_\beta,\rho_\beta\},
\label{eq:lio}
\ee
where $\{\; ,\; \}$ stands for anti--commutator. Conversely, the process
with values in pure states $\alpha,\psi_\alpha$ described in the previous
subsection is a unique one leading to (\ref{eq:lio}).
\end{theorem}
The equation (\ref{eq:lio}) describes time evolution of statistical states
of the total, classical+quantum, system. Sometimes, in special cases,
it is possible to sum up over $\beta$ to obtain evolution equation for
the effective statistical state of the quantum system alone. For this
being possible first of all the Hilbert spaces ${\cal H}_\beta$
must be identical. Then we can set $\rho=\sum_\beta \rho_\beta .$
Also, we must have the same Hamiltonian, and the same
$\Lambda$ in each \lq channel\rq\ : $H_\beta\equiv H, \Lambda_\beta\equiv
\Lambda$, moreover we must have special property that
$\sum_\beta \sum_\alpha g_{\beta\gamma}\rho_\gamma g_{\beta\gamma}^\star =
\sum_i V_\i \rho V_i^\star$ for some family of operators
$V_i$ which result from summing up subfamilies of operators $g_{\beta\gamma}$.
Only in that case we obtain Liouville's evolution equation
for $\rho$:
\be
{\dot \rho}=-i[H,\rho]+\sum_i V_i \rho V_i\star - {1\over2} \{\Lambda,\rho\},
\label{eq:lioq}
\ee
with $\Lambda=\sum_i V_i^\star V_i$. But even if this is the case, the
information lost is unrecoverable: there are always infinitely many
processes in the space of pure states of the quantum system that lead
to the same quantum master equation (\ref{eq:lioq}).
Even if equations (\ref{eq:lio}) and (\ref{eq:lioq})
look similar in form, there is an abyss of information loss that separates
their contents.
\section{Examples}
\subsection{Particle detector}
We consider the simplest case: that of a two--state classical system. We
call its two states "on" and "off". Its
action is simple: if it is off, then it will stay off forever. If it
is on, then it can detect a nearby particle and go off. Later on we will
specialize to detection
of particle presence at a given location in space.
For a while let us be
general and assume that we have two Hilbert spaces ${\cal H}_{off},
{\cal H}_{on}$
and two Hamiltonians $H_{off}, H_{on}$. We also have time dependent family of
operators $g_t :{\cal H}_{on}\rightarrow {\cal H_{off}}$ and let us denote
$\Lambda_t=g_t g_t^\star :{\cal H}_{on}\rightarrow{\cal H}_{on}$.
According to
the theory presented in the previous section, with $g_{off,on}=
g_t$, $g_{on,off}=0$,
the master equation for the total system, i.e. for particle and detector,
reads:
\begin{eqnarray}
{\dot \rho}_{off}(t)&=&-i[H_{off},\rho_{off}(t)]+g_t \rho_{on}(t) g_t^\star
\nonumber\\
{\dot \rho}_{on}(t)&=&-i[H_{on},\rho_{on}(t)]-{1\over2}
\{\Lambda_t,\rho_{on}(t)\}.
\end{eqnarray}
Suppose at $t=0$ the detector is "on" and the particle state is
$\psi (0)\in {\cal H}_{on}$, with $\Vert \psi (0)\Vert=1.$
Then, according to the event generating algorithm
described in the previous section, probability of detection during
time interval $(0,t)$ is equal to $1-\Vert \exp (-iH_{on}t-{t\over2}
\Lambda_t )\; \psi (0)\Vert^2 .$

Let us now specialize and consider a detector of particle presence at
a location $a$ in space (of $n$ dimensions).
Our detector has a certain range of detection
and certain efficiency. We encode these detector characteristics
in a gaussian function:
\be
g(x)=\kappa^{1/2}({\alpha\over\pi})^{n/2} \exp (-\alpha x^2),
\ee
where $n$ stands for the number of space dimensions.\\
If the detector is moving in space along some trajectory $a(t)$, and if
the detector characteristics are constant in time and space, then we put:
$g_t (x)=g(x-a(t))$.
Let us suppose that the detector is off at $t=t_0$ and that the
particle wave function is $\psi_0(x)$. Then, according to the
algorithm described in the previous section, probability of detection
in the infinitesimal time interval $(t_0,t_0+\Delta t)$ equals
$\int g_{t_0}^2 (x) | \psi_0 (x) |^2 dx \cdot \Delta t$. In the limit
$\alpha\rightarrow\infty$, when $g^2_t (x)\rightarrow \kappa \delta (x-a(t))$
we get $\kappa |\psi_0 (a(t_0)) |^2 \cdot \Delta t$. Thus we recover the
usual Born interpretation, with the evident and necessary correction that
the probability of detection is proportional to the length of exposure
time of the detector. \\
That simple formula holds only for short exposure times. For a prolonged
detection the formula becomes more involved, mainly
because of non-unitary evolution due to presence of the detector. In that
case numerical simulation is necessary. To
get an idea of what happens let us consider a simplified case which
can be solved exactly. We consider the ultra--relativistic Hamiltonian
$H=-i d/dx $ in space of one dimension.
In that case the non-unitary evolution equation is easily solved:
\be
\psi (x,t) = e^{-{1\over2}\int_0^t \Lambda_s(x+s-t)} \psi(x-t,0).
\ee
In the limit $\alpha\rightarrow\infty$ when detector shrinks to a point,
and assuming that this point is fixed in space $a(t)=a$,
we obtain for the probability $p(t)$ of detecting the particle in the time
interval $(0,t)$:
\be
p(t)=(1-e^{-\kappa}) \int_{a-t}^a |\psi(x,0)|^2 dx.
\ee
Intuitively this result is very clear. Our Hamiltonian describes
a particle moving to the right with velocity $c=1$, the shape of the
wave packet is preserved. Then $p(t)$ is equal to the standard
quantum mechanical probability that
the particle at $t=0$ was in a region of space that guaranteed passing
the detector, multiplied by the detector efficiency factor - in our case
this factor is $1-e^{-\kappa}.$
\subsection{Fuzzy clock}
This example illustrates diversity of possible couplings between
a classical and a quantum system. In the model below no information
is transferred from the quantum system -- except that passing of fuzzy units
of time is marked. The example also shows that the standard continuous
unitary evolution of quantum mechanics can be approximated with an
arbitrary accuracy by a pure jump process. \\
Again, as in the subsection above we will start with a setting which
is more general than usual -- we will work with a family of Hilbert
spaces rather than with one fixed Hilbert space. Those readers that
like to have only one Hilbert space may think that all our ${\cal H}_i$
below are identical to some standard Hilbert space ${\cal H}$.

\noindent
{\bf Remark:}
The situation here is similar to that of a relativistic Dirac's equation.
There is a separate Hilbert space for each space--like hypersurface,
namely the Hilbert space of Cauchy data. There are different possibilities
to identify these Hilbert spaces - different coordinate systems used
by different observers will lead to different identifications. Similar
situation occurs in Galilei general relativistic quantum mechanics -
see \cite{cajamo95}

For the classical system we take the set of clock readings i.e. the set
$Z$ of integers $i$. For each
$i$ we have a Hilbert space ${\cal H}_i$. As we have already said before --
there is no Hamiltonian part in the evolution. Concerning the classical
events: the only events that
we admit are clock's ticks. To each event $i-1\rightarrow
i$ we associate operator $g_{i,i-1}={\sqrt \kappa}U_i$, where $U_i$
is an isometry from ${\cal H}_{i-1}$ to ${\cal H}_{i} .$ Thus $U_i^\star
U_i=I$ and our master equation (\ref{eq:lio}) reads:
\be
{\dot \rho}_i=U_i \rho_{i-1} U_i^\star -\kappa \rho_i .
\ee
The associated process is of the simplest possible kind: at random
times, distributed according to the Poisson law with a constant rate
$\kappa$, the quantum state vector changes:
\be
{\cal H}_{i-1}\ni\psi_{i-1}
\rightarrow \psi_i=U_i \psi_{i-1} \in {\cal H}_i ,
\ee
and the classical clock pointer advances by one $i\rightarrow i+1$. The
clock is fuzzy and its clicks are random. If we want to count time
more uniformly we must collect large number of such clocks. But that is
not the point here. The main point of this example is to illustrate
our thesis: {\em no dissipation -- no information}. Indeed, there is
no dissipation in the quantum system in this example. Quantum pure states
evolve into quantum pure states. At the same time we learn nothing useful by
observing the classical system. We just learn that time has passed. And this
passage of time brings no information whatsoever about the quantum state. The
clock rate is constant -- it is completely independent of the quantum state.
\vskip10pt
\noindent
{\bf Acknowledgements}\\
This paper was completed at RIMS, Kyoto University. One of us (AJ)
thanks to Japan Ministry of Higher Education for supporting his stay and
to prof. H. Araki for his kind hospitality.

\end{document}